# A Search for Transient, Monochromatic Light from the Galactic Plane

Geoffrey W. Marcy[1*] & Nathaniel K. Tellis[2]

[1] *Center for Space Laser Awareness, 3388 Petaluma Hill Rd, Santa Rosa, CA, 95404, USA*
[2] *RocketCDL*

## ABSTRACT

The Galactic Plane was searched for transient, monochromatic light at optical and near-IR wavelengths to detect pulses shorter than 1 sec. An objective-prism Schmidt telescope of 0.28-meter aperture and a CMOS camera were used to observe 973 square degrees, with 8864 exposures of 1-sec each, within a strip 2.1 deg wide along the Galactic Plane, from Galactic longitude -4 deg to +248 deg. All exposures were analyzed for transient, monochromatic sources using a "difference image" algorithm that yielded 11 candidate sources. All 11 sources were found to be associated with either astrophysical emission-line objects or aircraft with sub-second blinking lights. Our survey "rediscovered" many Wolf-Rayet stars, M dwarf flare stars, and planetary nebulae. It also identified an aircraft, of unknown type, that apparently had a nearly monochromatic lamp and a xenon lamp. This survey would have revealed optical and near-IR pulses having a power of ~180 GW (wavelength dependent) if emitted by a 10-meter aperture laser located 1 kiloparsec away. These non-detections of laser pulses from the Galactic Plane, including a 10-degree region toward the Galactic Centre, add to the non-detections from more than 5000 nearby stars. Indeed, all-sky surveys for emission-line objects (e.g., ionized gas, supernovae remnants, and active galactic nuclei) would have revealed lasers of a wide range of average brightness, wavelength, and cadence. The absence of beacons reveals more of a SETI desert, notably at the intensely surveyed optical and radio wavelengths.

Key Words: Transients, Extraterrestrial intelligence, Galaxy, Techniques: Spectroscopic

## 1   INTRODUCTION

Surveys of the sky with time resolution have serendipitously revealed unexpected time-variable, dynamical astrophysics. Historic examples include eclipsing binary stars, Cepheid and RR Lyrae pulsating stars, flare stars, cataclysmic variables, and supernovae. More recent examples include previously unknown classes of objects such as pulsars, x-ray binaries, kilonovae, gamma ray bursts, and fast radio bursts (FRB). Surveys at radio wavelengths have enjoyed a natural propensity at finding unexpected sub-second transients because of the necessity to record voltage every ~$10^{-6}$ s or ~$10^{-9}$ s (at MHz or GHz frequencies), leading to the unanticipated discoveries of pulsars and FRBs. These fortuitous successes motivate searching unexplored domains of time and wavelength.

Very few all-sky surveys at ultraviolet, optical, and IR wavelengths have been done with exposure times shorter than 20 s, and fewer still with exposures shorter than 1 s. Long exposure times cause sub-second flashes to be diluted relative to the background night sky brightness, making them less visibl. The LaserSETI program at the SETI Institute employs millisecond exposures to overcome such dilution (Gillum 2022). The recent development of astronomy-quality CMOS sensors offers sub-second readout times, small pixels of a few microns size, and large pixel arrays >50 megapixels, providing access to transient phenomena having sub-second time scales. Here we report a search for monochromatic, sub-second pulses of optical and near IR light, motivated both by the unexplored domain of time and wavelength and by a speculative model of interstellar communication.

The Milky Way Galaxy may contain spacecraft, communication relay stations, or home star systems that communicate by lasers (e.g., Bracewell 1960,1973; Freitas 1980; Maccone 2021; Gillon 2014, Hippke 2020, 2021abc; Gertz 2018, 2021, Gertz & Marcy 2022). Indeed, lasers are already widely used for communication by Earth-orbiting satellites because they offer narrow-beams for privacy, high bit rate, and minimal payload mass. Similarly, laser communication, or next-generation coherent quantum communication methods, may be used for interstellar communication (Schwartz and Townes 1961, Zuckerman 1985, Hippke 2018, 2021abc), perhaps using repeater nodes (Gertz & Marcy 2022).

Laser light may be identified in telescopes by its narrow range of wavelengths, hereafter "monochromatic" (e.g., Naderi et al. 2016, Su et al. 2014, Wang et al. 2020), regardless of the unknown duration or cadence of the pulses. Previously, we have searched for monochromatic light from more than 5000 individual stars of all masses, ages, and chemical compositions (O,B,A,F, G, K, and M) using high-resolution optical spectra. These extensive searches yielded no laser detections and no viable candidates (Reines & Marcy 2002; Tellis & Marcy 2017, Marcy 2021, Marcy et al. 2022, and Tellis 2022, private communication). These laser searches employed spectra of high resolution, $\lambda/\Delta\lambda > 60000$, in the wavelength range $\lambda = 3600$ to 9500 Å. The detection threshold of laser power was 50 kW to 10 MW, assuming a diffraction-limited laser emitter consisting of a benchmark 10-meter aperture. We also searched a 10x10 degree field at the Milky Way Centre, albeit at low spectral resolution, yielding no laser detections (Marcy et al. 2022). We also searched the solar gravitational lens focal points for nearby stars (Marcy et al. 2021). No monochromatic light was found, neither sub-second pulses nor continuous emission.

Searches for narrowband radio waves from technological entities also make use of the "monochromatic" nature of a signal to distinguish it from ordinary astrophysical sources and to promote the candidacy of technological signals (e.g. Isaacson et al. 2017, Enriquez et al. 2017, Price et al. 2020, Wlodarczyk-Sroka, Garrett, & Siemion 2020, Sheikh et al. 2021, Garrett & Simion 2022). There is no guarantee that interstellar communication will be nearly monochromatic or multi-bandpass. But that characteristic offers a search property that excludes the vast majority of false positives that are either astrophysical or terrestrial noise.

Without spectroscopic information, other searches for technological signals have been done by hunting for sub-second optical pulses, e.g., Wright et al. (2001); Howard et al. (2004); Stone et al. (2005); Howard et al. (2007), Hanna et al. (2009), Abeysekara et al. (2016), Villarroel et al. (2020, 2021). No definitive optical pulses were found, but some candidates emerged. Next generation searches for sub-second optical pulses are planned (Maire et al. 2020). Here we describe a search for sub-second pulses of monochromatic light along the Milky Way Plane using a special optical system designed to optimize this search.

## 2 OBJECT-PRISM OBSERVATIONS OF THE MILKY WAY PLANE

We used the objective prism Schmidt telescope operated by the Center for Space Laser Awareness and described in Marcy et al. (2021, 2022) and at www.spacelaserawareness.org . In brief, the telescope is a modified Schmidt design with aperture diameter 0.28 m and a 7-degree wedge prism to produce spectra of low resolution, R~100, of every point within the 2x3 deg field of view. The CMOS camera at the focal plane contains 9400 x 6600 pixels, each 3.7 microns with a quantum efficiency over 80% between 500 – 800 nm and lower QE extending to 370 nm and 950 nm. We operated with an exposure time of 1.0 sec and dead time of < 0.01 sec.

The system is optimized to detect monochromatic pulses of optical light (see Marcy et al. 2021, 2022) having pulse duration less than 1 sec. Monochromatic emission has the shape of a two-dimensional PSF, with a FWHM ~ 5.5 pixels, allowing efficient search algorithms. In contrast, direct hits by particles or gamma rays (cosmic rays) make "dots" sharper than the PSF, allowing discrimination. Another nemesis comes from reflections off satellites (Corbett et al. 2021, Nir et al. 2021) which exhibit the solar spectrum, obviously not monochromatic. The optical design is remarkably similar to the objective prism telescopes of the Harvard Observatory (Pickering 1912; Fleming 1917), albeit with 50x the QE.

We performed multiple performance tests of the QHY600 CMOS camera, finding read noise is ~2 photons (RMS), the dark noise <0.1 e$^-$/s per pixel, and the response is linear within 0.3% over a dynamic range of a few to 56,000 photo-electrons. We operate with modest thermoelectric cooling to -20C. The properties of the QHY600M are comparable to CCDs (Gill et al. 2022, Betoule et al. 2022), but offer frame rates up to 30 fps. We can detect light pulses having sub-second duration with minimal contamination from the background "noise" of stars, galaxies, and sky. From our observing station at Taylor Mountain in California, the sky produces ~40 photons/pixel during 1 sec coming mostly from Santa Rosa city lights.

Figure 1 shows a sum of 10 images obtained with the objective prism telescope system and QHY600 CMOS camera. This image is centred at Galactic Longitude 38 deg and Galactic Latitude 0 deg. The image shows hundreds of stellar spectra oriented vertically, each spanning wavelengths 380 – 950 nm spread over 1200 pixels, with long wavelengths downward and North up. In the 1 sec exposures, the faintest stars have Vmag=13 with signal-to-noise ratios of ~10 per pixel. Stars brighter than Vmag = 2.5 saturate the sensor with >56000 photons/pixel. A monochromatic, spatially unresolved point source would appear as a two-dimensional "dot" with a PSF shape. Sub-second monochromatic pulses would appear in only one image as a PSF-shape "dot" within a sequence of images. This survey searches for cadences of at least one pulse every 10 minutes, the duration of the observing sequence of a given field. Continuous sources and high cadence sources, >1 Hz, would appear in all images in a sequence of 600 images. We judge the PSF by the width of the spectra in the spatial direction that is dominated by both seeing and optical imperfections in the prism, yielding a PSF width of typically 6 to 8 arcseconds corresponding to a FWHM 5 to 6 pixels.

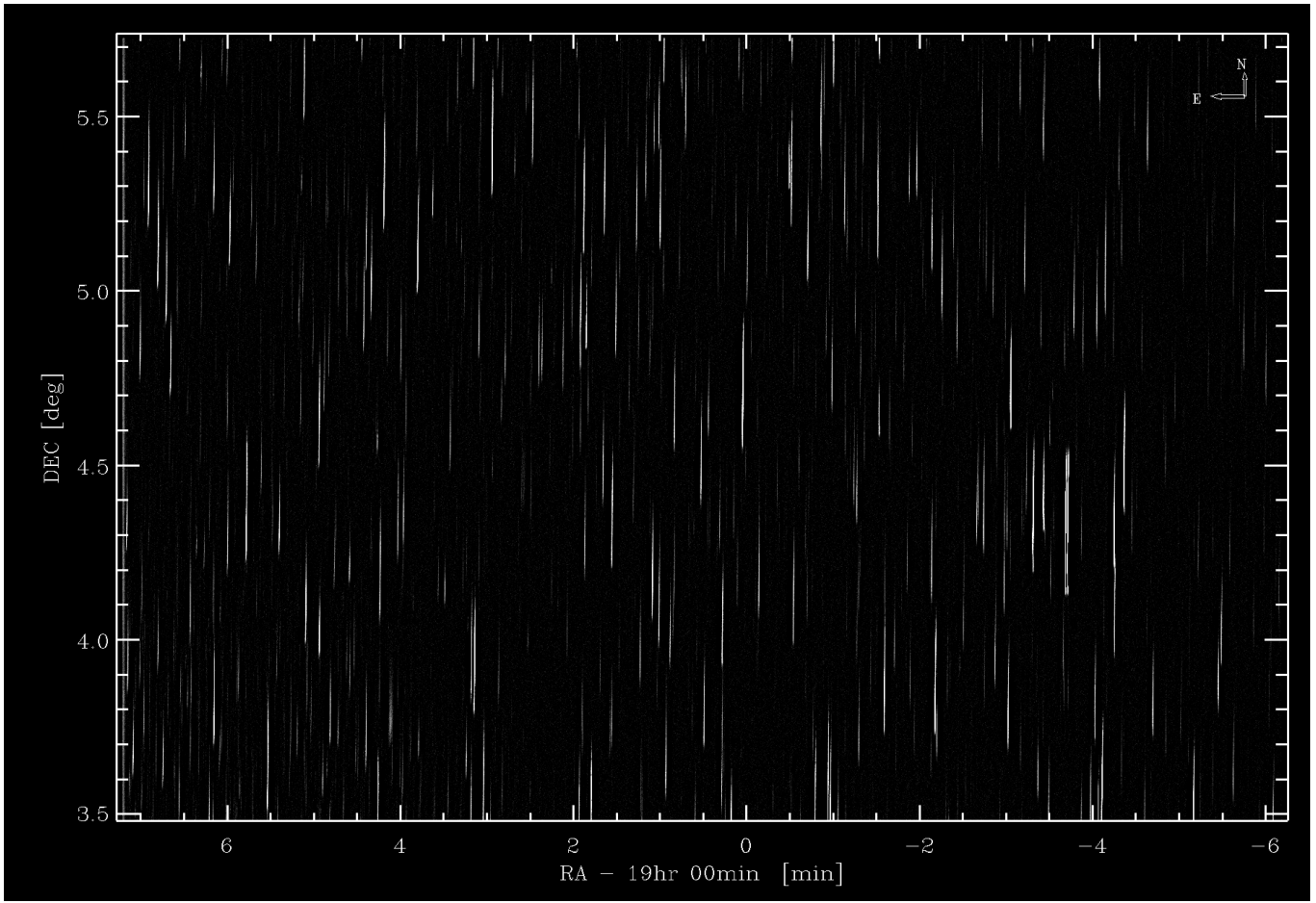

*Figure 1. The sum of 10 images (1 sec each) from the objective prism system with a field of view of 3.1 x 2.1 deg, 9500 x 6300 pixels, each subtending 1.3 arcsec on the sky. The stellar spectra span wavelengths 370 – 950 nm with longer wavelengths downward and north up. This image is centred at Galactic longitude 38 deg, at RA=19h 00m, DEC=+4º 30'. The stellar spectra come from stars of magnitude 8 – 14. Monochromatic emission would appear as a PSF-shape "dot".*

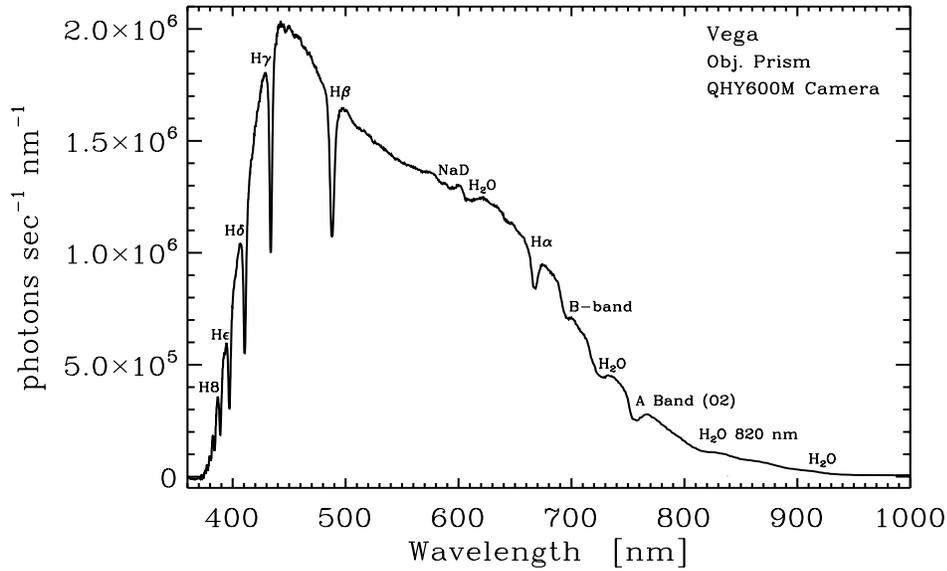

*Figure 2. Spectrophotometry of Vega vs wavelength obtained with the objective prism telescope and CMOS camera. The spectrum establishes the spectrophotometric sensitivity, wavelength scale, and spectral resolution, R~100, of all spectra that have a length 1200 pixels. Prominent absorption lines in Vega and strong atmospheric lines are labelled. Exposure time was 0.5 sec.*

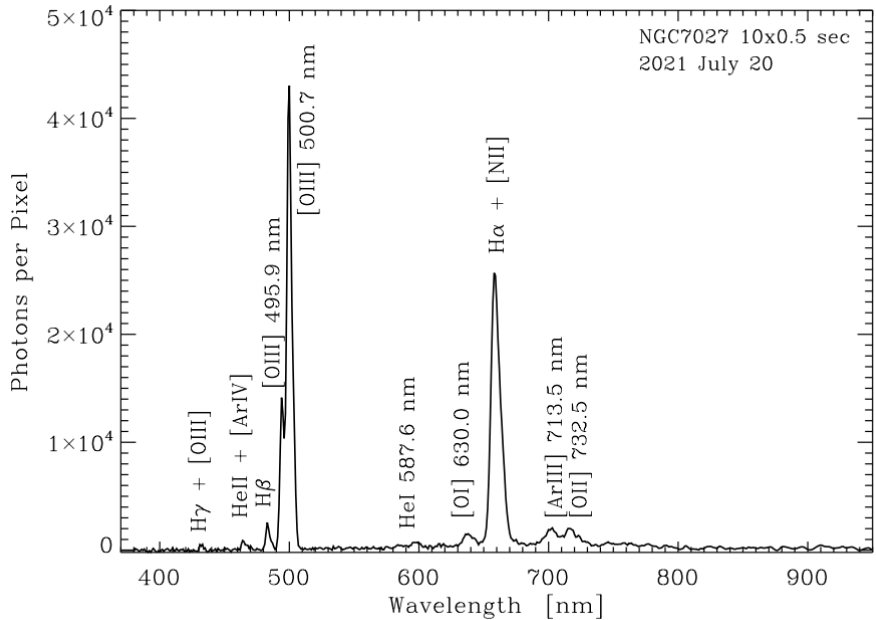

*Figure 3. A spectrum of the planetary nebula NGC7027 with the objective prism system and 5 sec exposure total. It shows emission lines common from ionized gas. The wavelength scale is based on a spectrum of Vega, with a zero point set by H$\alpha$ here. Laser emission is easily distinguished from known astrophysical sources by their pattern of known emission lines.*

Figures 2 and 3 show spectrophotometry of Vega and NGC7027 obtained with the objective prism system. The reduction to one dimensional spectra was accomplished by a simple summing of the photons along the spatial width at each wavelength. The spectrum of Vega (0.5 s exposure) shows the Balmer lines up to H11, along with telluric lines (Fig.2). The wavelength calibration was done with a 7$^{th}$ order polynomial fit to 14 pixel positions and the corresponding wavelengths in that Vega spectrum. The prism creates a highly nonlinear wavelength dispersion. The resulting Vega spectrum in Fig. 2 shows that stellar spectra can be classified and that emission-line spectra can be identified, to distinguish them from non-astrophysical sources. The spectrophotometry of Vega in Figure 2 is given in photons per nm per sec detected with our objective prism system, allowing this spectrophotometry to map magnitude to photons per nm per sec of other sources. The monotonic decrease in photons detected for wavelengths shortward of 440 nm is due to decreasing quantum efficiency of the CMOS sensor.

The spectral and spatial resolutions are set by the PSF that has FWHM ~5.5 pixels, dominated by seeing and optical aberrations in the prism. The spectrum of NGC7027 (Figure 3) shows the usual emission lines from ionized gas at 10000K (e.g. Zhang & Li 2003). The two [OIII] lines at 495.9 and 500.7 nm are barely resolved. This modest spectral resolution, 2 to 10 nm from 380 to 950 nm, allows the identification of stars, galaxies, ionized gas, asteroids, aircraft, and satellites, to distinguish them from less common phenomena.

Reflected sunlight and glints from orbiting satellites are easily identified by their solar spectrum. This allows light pulses from Earth-orbiting satellites and discarded rocket boosters to be immediately distinguished from extraterrestrial laser pulses. The spectroscopic information allows instant discrimination of both astrophysical and terrestrial sources from monochromatic extraterrestrial sources. However, laser pulses from human-made satellites could be indistinguishable from extraterrestrial laser pulses. Satellite-born laser pulses that are sufficiently brief to avoid detecting an orbital "streak" on the image, i.e. less than ~1 arcsecond, can masquerade as extraterrestrial laser pulses.

Figure 4 shows the RA and DEC of the fields we observed, each field having of angular size 3.2 x 2.1 deg located along the Galactic Plane. Also shown on Figure 4 is the 14 x 10 deg region we previously observed near the Galactic Centre, as labelled. The total coverage is 973 square degrees along the Galactic Plane. Missing is the Galactic Plane south of 34 deg, inaccessible to our observatory in Northern California. The field toward the Galactic Anti-Centre, at RA = 5h 45m 37s DEC=+28 56', was observed four times, as that direction is special. At the Anti-Centre, laser guide stars or communication lasers may be pointed toward us, but actually intended for the Galactic Centre.

Each field was observed with 600 consecutive 1-sec exposures. This survey detects arrival of at least one photon pulse every 10 minutes. The fields overlap to provide both complete coverage of the region and security against algorithmic or optical poverty at the edges of the field such as from poor background assessment or vignetting.

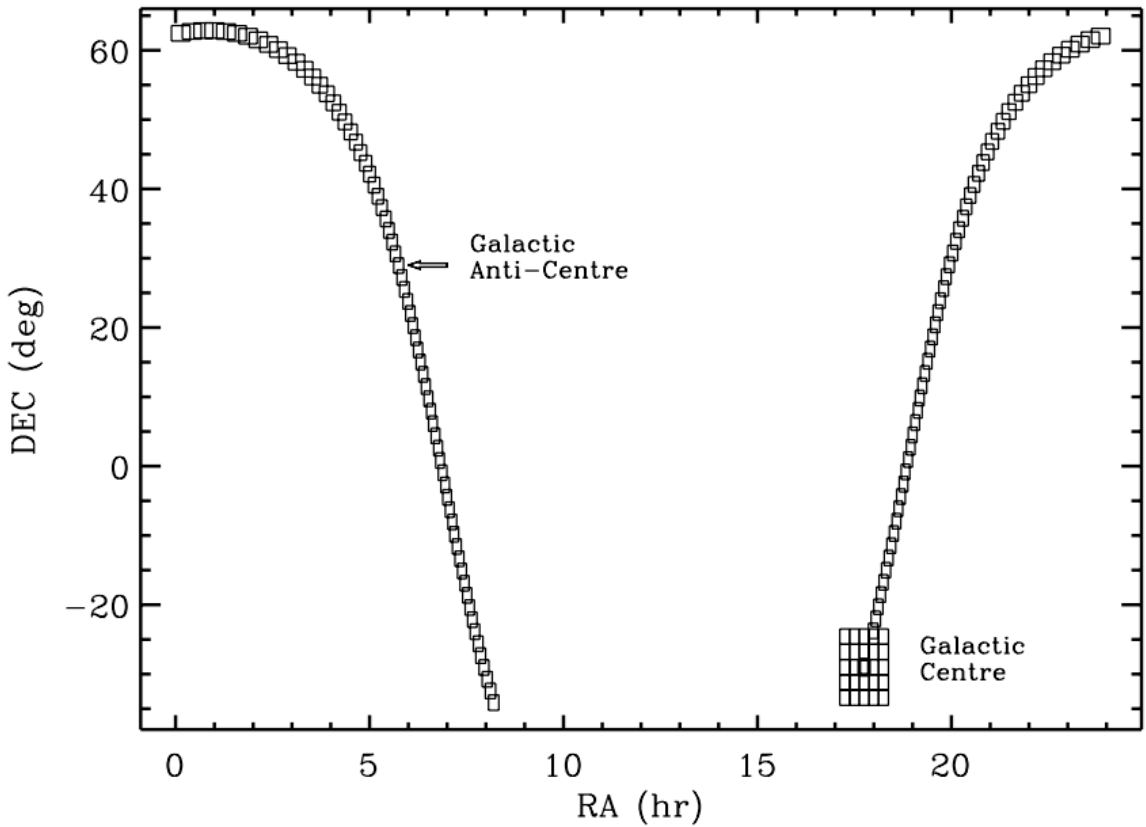

Figure 4. The 124 fields, each 3.2x2.1 deg, observed along the Galactic Plane in this objective prism survey. Also shown are the previously observed fields near the Galactic Centre, for a total of 973 square degrees surveyed. Each field was observed

*with 600 consecutive 1-sec exposures, giving time-resolved spectra of all points in the sky to reveal sub-second or continuous monochromatic optical pulses.*

## 3   THE DIFFERENCE-IMAGE ALGORITHM

We search for monochromatic emission that appears as a transient PSF-shape "dot" in the images. We employ a difference-image technique described in Marcy et al. (2022), similar to the difference algorithm in Vasilyev et al. 2022. In brief, the algorithm operates on a set of 600 exposures, each 1 s, for a specific 2x3 deg field. The algorithm takes the average of six "bookend" images (three prior and three subsequent) surrounding a given "target" image and subtracts the average bookend image from the target image to yield the "difference image", having pixel values near zero. Residuals are due to Poisson noise of the arrival of photons and to the variations in atmospheric "seeing" from image to image that compromises the subtraction of stellar spectra. We suppress these residuals by performing a 50-pixel boxcar smoothing of the difference image along the direction of dispersion of the spectra, and we subtract that smoothed version from the original difference image (see Marcy et al. 2022). This process subtracts the residual continuum of each star spectrum. Narrow emission lines in the target image that are not in the bookend images will persist in the difference.

This difference-image algorithm yields any monochromatic point sources that were present in each image but not present (or only weakly present) in the average of the six "bookend" images. Examples of this process are shown in Figure 5 of Marcy et al. (2022), and we used the same algorithm here. Each target difference-image (9500x6300 pixels) was examined blindly by this algorithm, yielding PSF-like candidates.

Emission-line sources may be coincident with stellar spectra or they may be located in between them, and the difference-image algorithm suppresses light from both stars and the sky. The algorithm further demands that the candidate point sources must have a 2D shape consistent with the instantaneous point spread function (PSF), as measured by the spatial profile of the stellar spectra determined by cross-correlation. For each exposure, the algorithm measures the FWHM of the spatial profile of stellar spectra, commonly 5 to 6 pixels (6 to 8 arcseconds), caused by seeing and optical aberrations in the prism. We note that "cosmic-ray" particles that hit the CMOS sensor are immediately rejected, as they affect only a few neighboring pixels, inconsistent with the smooth Gaussian-like shape of the PSF with 5.5 pixel FWHM.

The algorithm is designed to detect sub-second monochromatic pulses. However, if the cadence of light pulses is more frequent than 1 pulse per second, the image-difference algorithm will be compromised in detecting them because the pulses appear in both the target image and the bookend reference images. If the pulse amplitudes are not constant or if the seeing changes over a period of 7 s, the pulses may still be detected. Cadences slower than ~1 pulse per second will yield individual frames containing the point source and neighboring ones that do not, making the pulse detectable with the difference algorithm.

Somewhat surprisingly, sources of monochromatic light that are *constant in time* are usually detected by the difference-image algorithm despite appearing in both the target and bookend images. The seeing changes on sub-second time scales causes the captured number of photons to vary by more than 10% from image to image. The intensity of emission lines from astrophysical objects such as flare stars, planetary nebulae, and Wolf-Rayet stars varied in apparent brightness by 10% - 20% (RMS) among the 1 s exposures. The emission lines are detected by the difference-image algorithm as if they were transient sources, even though they are actually steady. Thus, the algorithm actually detects monochromatic point sources no matter if they last less than 1 s or are continuous in time.

## 4   DETECTED MONOCHROMATIC CANDIDATES

We executed the difference-image algorithm to all 124 fields and their 600 exposures per field along the Galactic Plane region (Figure 4), yielding 11 monochromatic objects of interest, each requiring visual inspection and assessment. We describe the monochromatic candidates here.

### 4.1 Moving Multiple Flashes: Probably Aircraft

On 2022 Jan 24, we obtained 600 exposures, 1 s each, at Galactic longitude 92 deg for which the automated difference-image algorithm gave an alert of several monochromatic sources. Upon inspection, the object clearly had several components, as seen in Figure 5. That figure shows the seven consecutive raw images, each panel showing the full frame image, 2.1 x 3.2 deg, with north to the left and longer wavelengths to the right. A few hundred stellar spectra are apparent, of V magnitudes 6 to 13, fixed in each image. The field has coordinates RA=21h 20m 30s +49º 46', in the northwest region of the sky at an altitude 40 deg., and the angular velocity of the object is 0.55 deg/s.

An odd-shaped object first appears (barely) in the upper left of the first frame, and it moves down and to the right each successive image during the full 7 s. Successive locations along a diagonal path, from upper left to lower right, represents time.

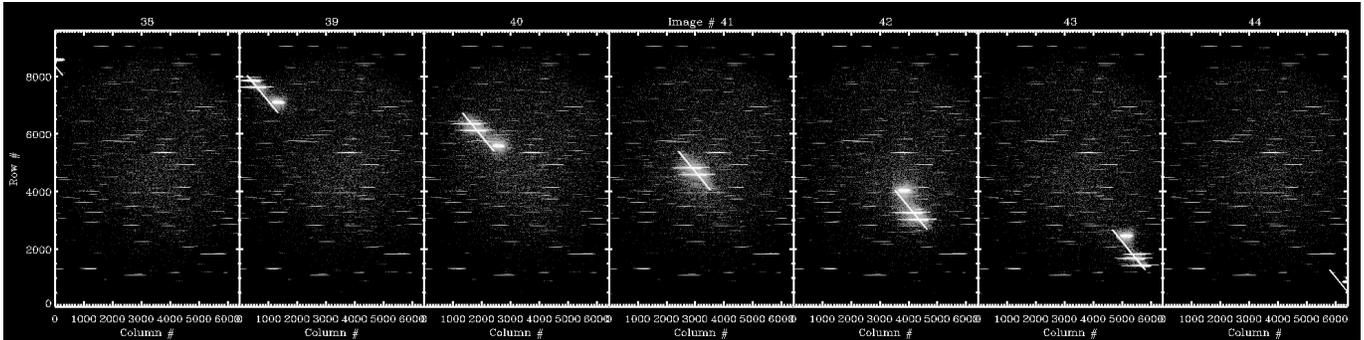

Figure 5. A sequence of seven consecutive full frames, each a 1.0 s exposure, at Galactic longitude 92 deg observed on 2022 Jan 24 at 2:33 UT. An object enters the field of view at upper left and moves down and to the right each second. During 1.0 s, several flashes occur having a broad spectrum, 1000 pixels across left-right, and one flash (the brightest) occurs having a smaller range of wavelengths. The diagonal line in each frame represents a light that is shining continuously during the full 7 s. It is apparently a nearly monochromatic light because it exhibits very little extent in the wavelength direction (left-right).

A magnified view of the third frame is shown in Figure 6, in which the stellar spectra were subtracted using the previous and next images, leaving only the moving object. Figures 5 and 6 show that the object travelled from upper left to lower right during each second, including a continuous diagonal line caused by a light source that stayed on during the entire 7 seconds of the sequence. Remarkably, this "diagonal light source" has a wavelength extent, left to right length, that is only ~60 pixels FWHM, corresponding to a wavelength spread of ~22 nm. This light is nearly monochromatic. We do not know its central wavelength, nor how this continuous light was produced.

Figures 5 and 6 also show at least six flashes of light that happened during the 1 second exposure. Each flash caused a long horizontal line showing that the light in the flashes consisted of the full optical wavelengths, 380 – 950 nm, similar to the background stars. The horizontal lines are narrow, only ~20 pixels wide along the diagonal (time) direction, corresponding to ~1% of the full diagonal length of travel during 1 s. This shows that each flash lasted ~0.010 s. The spectra of all 6 flashes show spectral structure at the far right end (longest, near-IR, wavelengths), consisting of at least to emission bumps.

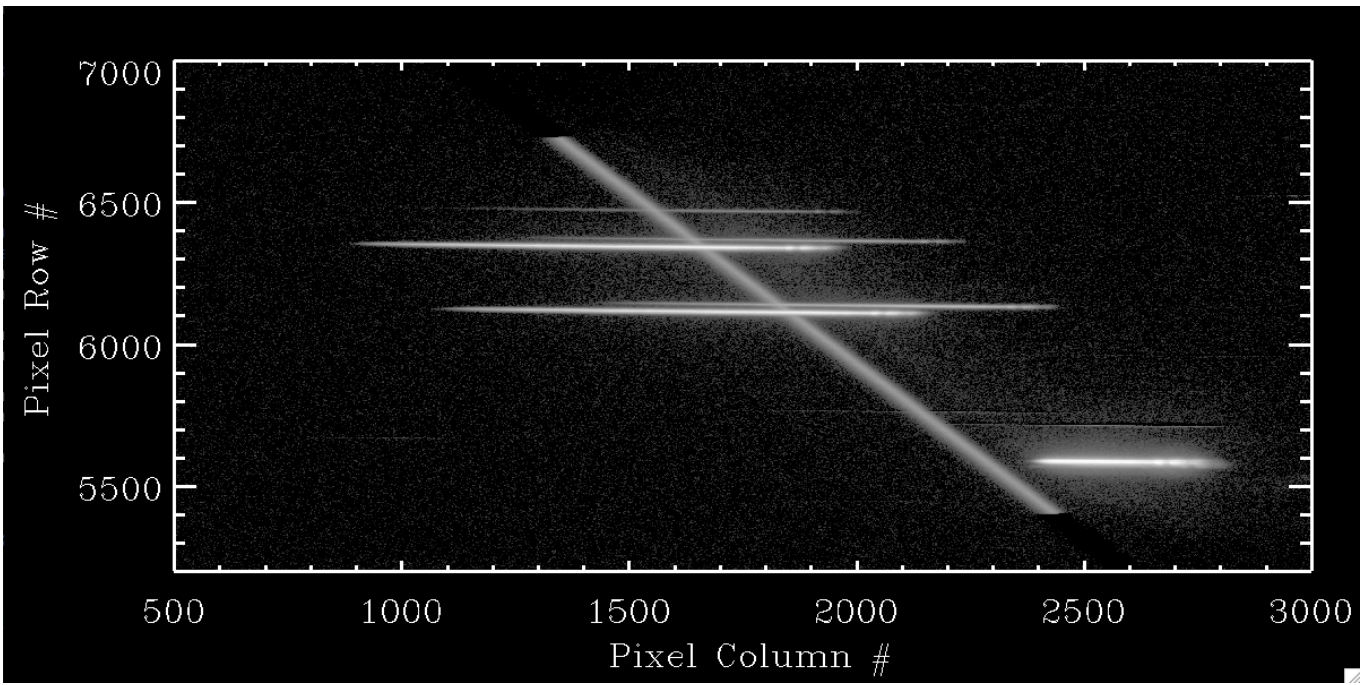

Figure 6. The magnified view of the laser candidate in Figure 5, the 3rd image. Background stars were subtracted using adjacent images. The diagonal streak is caused by motion of a light source from upper left to lower right and shining during the full one second. Its narrow width, left-right, shows it was nearly monochromatic, still not understood. Each of the six flashes exhibits a broad range of wavelengths (left-right extent), with emission lines at the longest wavelengths (at the right end of each flash, consistent with xenon lamps. The brightest flash (bottom right) is only 400 pixels long, indicating a short wavelength range, but the same emission lines.

The brightest horizontal line located at bottom right in Fig. 6 happened at a time ~0.9 sec during the 1.0 s exposure, judging from its location along the diagonal path. It contains only 1/3 of the full range of wavelengths. The emission structure at its far right of its spectrum resembles the emission structure at the far right of the other flashes. Apparently, those wavelengths are at the far red and infrared end of the spectrum. That brightest flash at bottom right also has a duration of ~0.01 s. One other attribute of all of the lights is that they continued to illuminate the telescope during the full 7 s. This suggests that all the light sources emitted a beam that was broad enough in solid angle to keep the telescope bathed in light during the full 7 s of the object's motion across 4 deg of the sky. To accomplish this steady bathing, the beams of light must have been many degrees across at least, and perhaps nearly isotropic.

We wonder what combination of light sources can create the multi-component images in Figures 5 and 6, and what type of object they are attached to. A clue comes from spectral features in the full-length spectra. At the long wavelength end of each spectrum (far right) is clearly some structure, and perhaps broad emission lines. We extracted those spectra from the raw image by simply adding 10 rows along the full length of the spectrum, one of which is shown in the left panel of Figure 7. The spectrum shows two broad, strong emission lines in the near-IR, at wavelengths ~830 nm and ~900 nm, along with some weaker emission lines at 480 nm and ~540 nm. Laboratory spectra of xenon displays its two strongest lines at 850 and 910 nm, in good agreement with the two strongest seen here in Figure 7 (Povrozin & Barbieri 2016, and https://mmrc.caltech.edu/Stark/Xe%20lamp%20spectra.pdf). The third strongest emission feature fin the observed object is a pair of closely spaced lines at ~470 nm (see Figure 7) that also agrees with the lines seen in lab xenon. The spectrum in Figure 7 also has a close pair of lines at 540 nm, which is not in the lab xenon spectrum. This may indicate another gas, besides xenon, in the lamp, such as mercury. Large airplanes typically have at least eight different external lights, usually composed of xenon gas, each having different color filters and beam directions, some of which flash at intervals near 1 second. The angular speed of the object, 0.55 deg s$^{-1}$ is consistent with that of aircraft by common experience, i.e., a Moon diameter per second.

Aircraft xenon lamps often have filters to produce "green" and "red" lights. LED lights are now also being used on aircraft. We wonder if an LED coupled with a narrow band filter could produce the "diagonal" line that was on continuously with its narrow wavelength range. LIDAR seems unlikely, as the lasers usually operate at 1064 nm, or frequency doubled to 530nm, both having narrow wavelength ranges of < 1 nm. The observed wavelength range of 22 nm is inconsistent with such lasers. Alternatively, a xenon lamp that was continuously on, but covered by a narrowband filter, could produce the persistent diagonal line. There is also the possibility of reflected light, laser or otherwise, off the aircraft.

In any case, the most likely explanation for this "monochromatic object of interest" is an aircraft with multiple, flashing lights with different filters, with one lamp shining continuously. Indeed, the northwest region of the sky is the direction of a small airport, Charles M. Schulz Airport, 18 km away. We don't know what causes the continuous "diagonal" light.

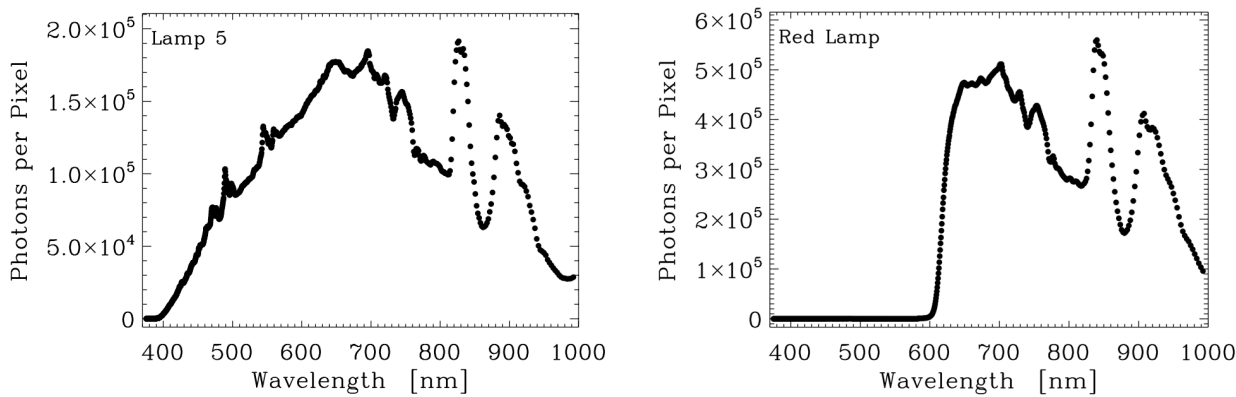

*Figure 7. Photons per pixel vs. wavelength of the two bright flashes in rows 6120 and 5580 in Figure 6. Both spectra show emission lines at 830 and 910 nm, which match the known strongest emission lines in Xenon gas, commonly used in airplane lamps. The weak emission lines at 475 nm here also appear in laboratory spectra of Xenon. The spectrum at right is missing light shortward of 600 nm, undoubtedly caused by a filter that transmits only light longward of 600 nm, making a "red" light. Thus, the lamps are made of xenon gas, as is commonly used on aircraft.*

On 30 Nov 2021, observations at Galactic longitude 46 deg revealed a new object of interest, detected by the automated difference-image algorithm. Figure 8 shows a sequence of seven full frames, each 1 second duration. As with the previous candidate (Figures 5, 6, 7), there are several flashes lasting less than 0.05 sec, several being full spectrum and one being red and bright. All of them contain the strong emission lines in the near-IR indicating xenon lamps again. A light with a continuous spectrum is also apparent, again lit continuously, seen as the faint diagonal streak. The angular speed is again ~1 deg/sec. We presume this object is also an aircraft.

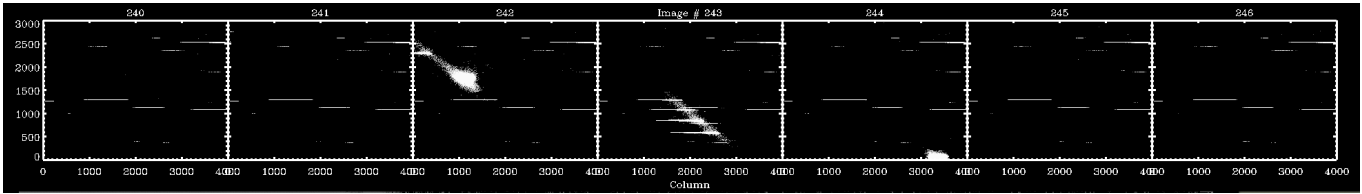

*Figure 8. A transient found at Galactic longitude 46, seen entering the field in the third frame and exiting in the 5th frame. The full spectrum flashes of light, and diagonal trajectory, are consistent with the xenon lights on an aircraft*

.

## 4.2 Wolf-Rayet Stars

At Galactic longitude 8 deg, the automated difference-image code identified a transient emission line among the 600 1-s exposures. Extraction of the raw image along the full spectrum revealed other emission lines, shown in Figure 9. We accomplished the wavelength calibration by employing, blindly, the calibration of wavelength vs pixel obtained from the spectrum of Vega used in Figure 2. The zero-point of a wavelength scale was set by using the pixel at the far infrared end of the spectrum in the raw image, deemed to be a wavelength of ~950 nm. The uncertainty of that zero-point is ~50 nm, due to the gradual dimming at the end of the spectrum. This approximate wavelength scale allowed the emission lines to be identified, within 50 nm, as shown in Figure 9.

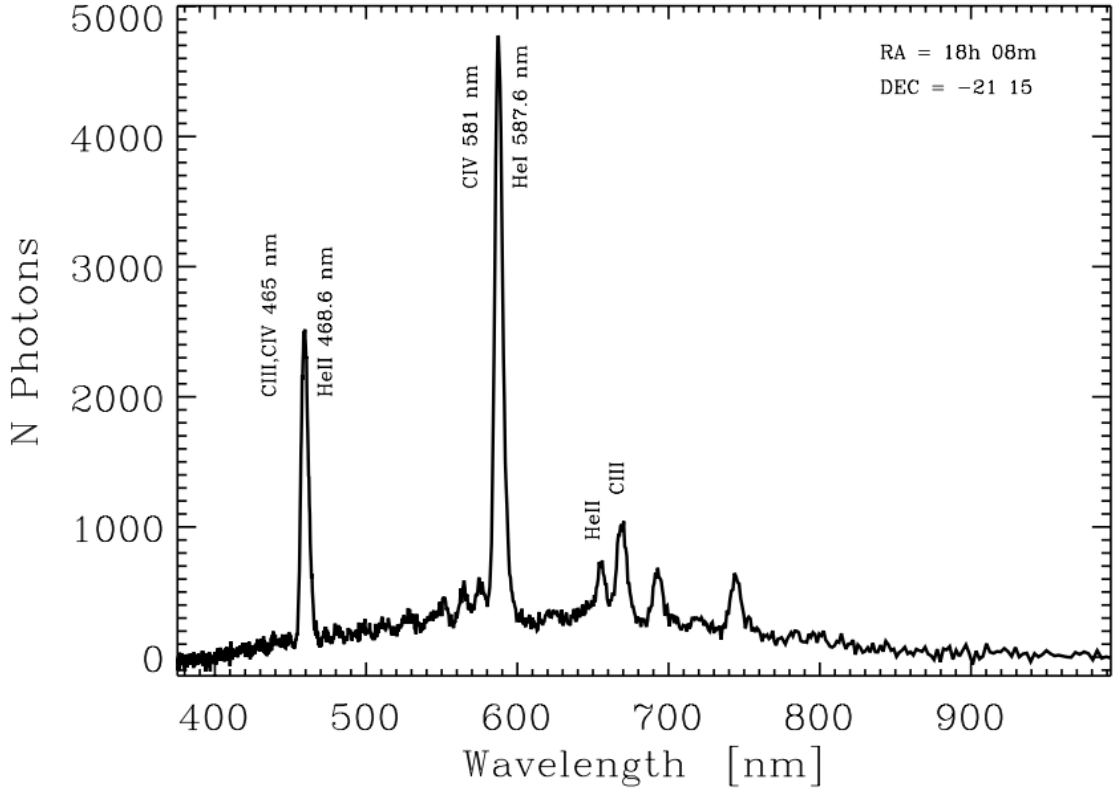

*Figure 9. Spectrum of a changing emission line at wavelength 585 nm, identified by the automated difference-image algorithm, in the field at Galactic longitude 8 deg. The extracted spectrum shows other emission lines, consistent with a Wolf-Rayet star of type WC5. The coordinates from our image (upper right) suggest this is the known Wolf-Rayet star, "WR111".*

That approximate wavelength scale showed the similarity of the spectrum with a standard Wolf-Rayet star of type WC5, based on the catalog Wolf-Rayet spectra given at: https://lweb.cfa.harvard.edu/~pberlind/atlas/htmls/wrstars.html and on the classification by Smith et al. (1996) and Crowther et al. (1995). Using the lab wavelengths of identified lines, we refined the zero-point of the wavelength calibration to yield the spectrum shown in Figure 9. We performed astrometry of our image, yielding coordinates, RA = 18h 08m and DEC=-21d 15', at which exists the known Wolf-Rayet star, WR 111=HD 165763 (type WC5). This obviously removes this candidate from further consideration as non-astrophysical.

A Galactic longitude 74 deg, the automated code revealed another emission line that varied in brightness during the *600 1-sec*. It met the criteria of a pulsing laser candidate in 28 of the exposures. Figure 10 shows the full spectrum, revealing the emission line at 465nm, and also several other emission lines. The line at 465 nm appears to vary in intensity because of seeing changes, easily verified by the widths in the spatial direction of the spectra. The pattern of emission lines matches that of Wolf-Rayet stars of type WC8, with coordinates approximately RA=20h 15m and DEC = +36d 38', with a brightness approximately V = 9, with a possible identification as HD 192641, a WC7 Wolf-Rayet star. Thus, there is no support for a non-astrophysical interpretation, laser or otherwise.

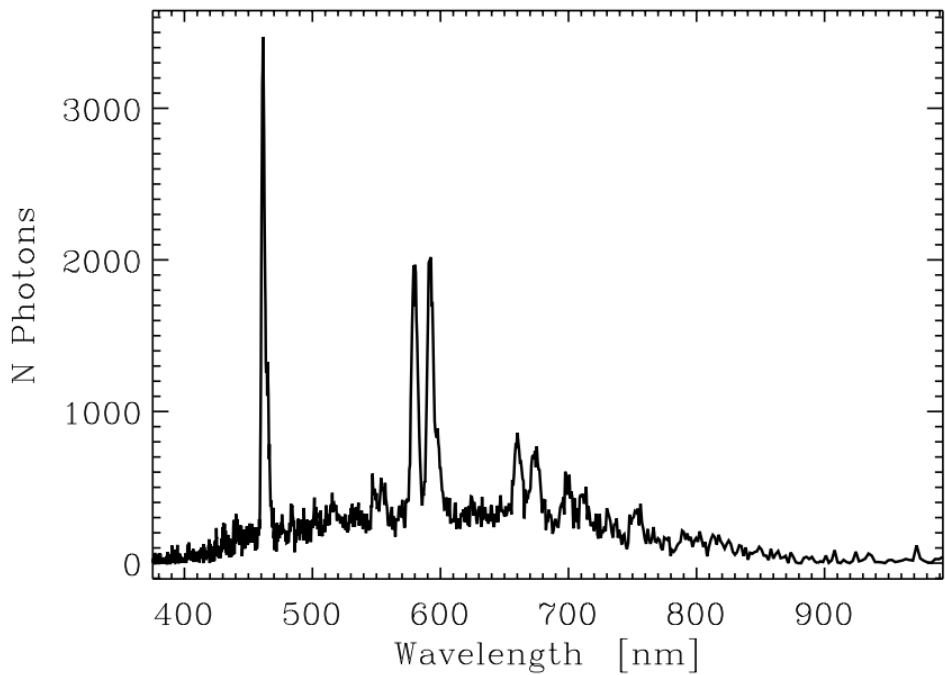

*Figure 10. An emission candidate detected automatically by the intensity variations of the emission line at 465nm due to changes in seeing. This is apparently a WC8 Wolf-Rayet Star of magnitude V~9, at approximate coordinates, RA = 20h 15m , DEC=+36d 38'. We rule out extraterrestrial technology.*

In the field at Galactic longitude 76 deg, the automated search triggered on another candidate transient emission line, shown in Figure 11. The spectrum shows it to be a WC7 Wolf-Rayet star, and the coordinates show that it is HD 192641, an 8[th] mag WC7 Wolf-Rayet star. The automated search for transients was triggered by seeing variations.

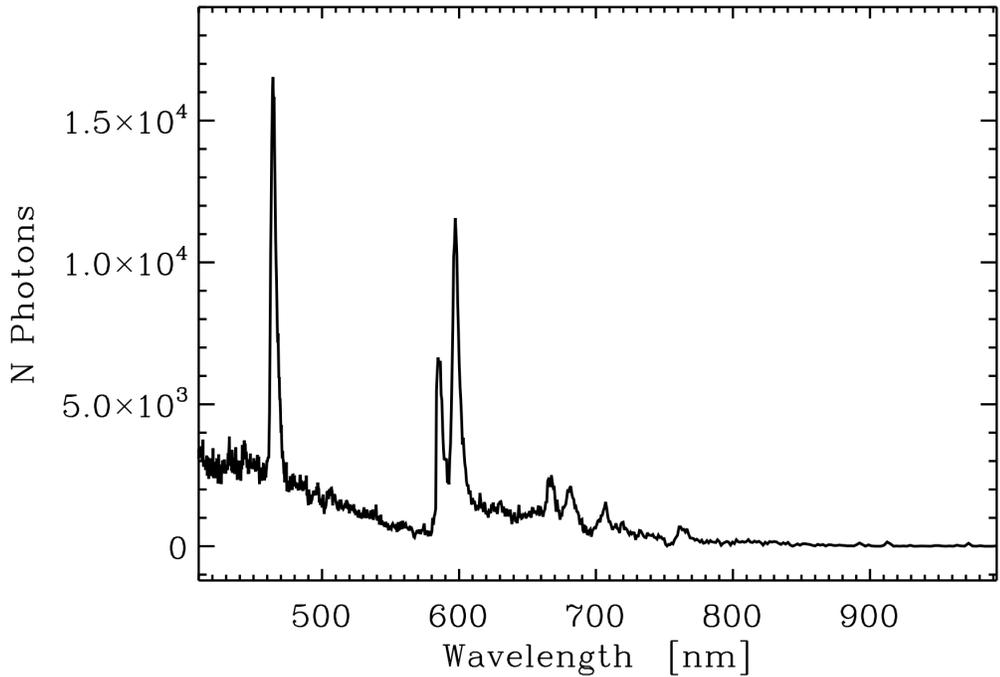

*Figure 11. The automated search for transient emission lines discovered this candidate, which is clearly a WC7 Wolf-Rayet star. This spectrum results from adding 20 1-sec exposures to improve the signal-to-noise ratio. Seeing changes caused the most intense emission line at HeI 465 nm line to momentarily brighten, triggering the alert. It is likely HD 192641, a WC7 Wolf-Rayet star.*

4.3  M-Type Stars

At Galactic longitude 38, our automated code revealed an apparent, changing emission line at a wavelength of 720 nm.  A plot of the full spectrum, shown in Figure 12 (top), shows a spectral energy distribution that is clearly an M dwarf, with the characteristic TiO absorption bands at the red and near-IR wavelengths (Leggett et al. 2000).  There is a naturally occurring peak at 720 nm.  Examination of a sequence of seven images, shown in Figure 12 (bottom), shows that the momentary improved seeing conspired to yield an apparent increase in the peak intensity, fooling the code into sensing an emerging emission line.  We see no evidence of a PSF-like line, and instead relegate this candidate to the common occurrence of momentarily improved seeing at a wavelength with natural high intensity.

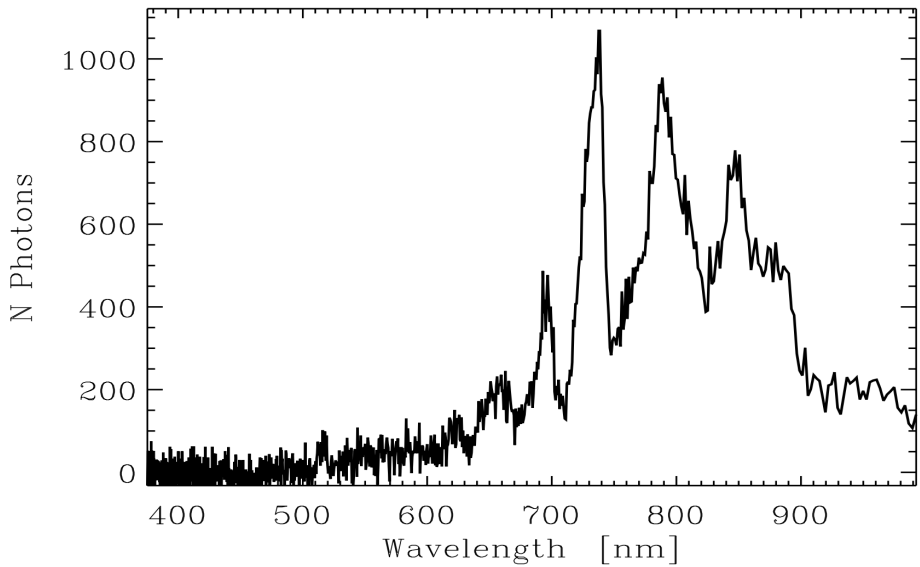

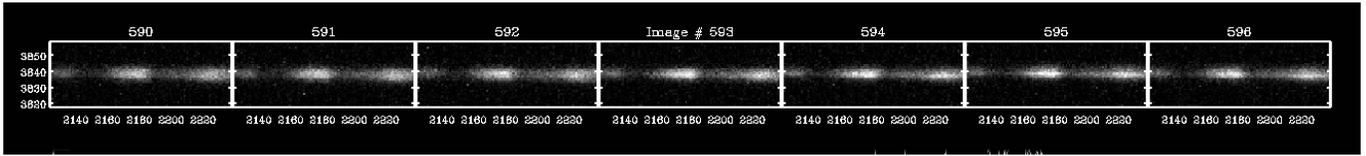

*Figure 12. A candidate emission line at a wavelength ~738 nm (column 2181) at Galactic Longitude 38 deg.. Top panel: the spectrum of the object, clearly an M dwarf (M4 to M5) with an apparent emission line at 738 nm. Bottom Panel: the seven consecutive raw images zoomed on the emission. The apparent intensity increase of the emission is just due to momentary improved seeing at a peak of the spectral energy distribution, between TiO absorption bands.*

At Galactic Longitude 72 deg on 2021 Dec 02, the automated code identified a monochromatic brightening at wavelength 827 nm (column 6167), in a star with an M4 spectrum. Figure 13 shows the extracted spectrum vs column #, both for the single exposure (left panel) that yielded the candidate and for the sum of 20 exposures (right panel) that gives an average spectrum over 20 sec. The candidate identified by the difference-image algorithm is located at a persistent peak in the spectrum of the M4 star. On that one exposure there was a momentary enhancement of the intensity that is obviously consistent with the noise level in that single exposure, thus making the apparent transient emission line to be likely noise. Indeed, examination of the raw image shown in Figure 14 shows that the enhanced emission was concentrated within 4 pixels, which is inconsistent with the PSF that has FWHM ~5.5 pixels. Thus, we suggest that this candidate monochromatic emission is simply noise. Follow-up spectroscopy of this M star may be warranted.

Unfortunately, the identity of the M star in Figures 13 and 14 remains a mystery. The coordinates from our images are approximately RA= 20h 13m 10s, DEC=+33d 07' (eq 2000, epoch 2021). From our image, its magnitude is Rmag ~11, where two plausible stars reside, HD331958 and TYC 2675-1608-1, neither of which has high quality characterization in the literature. HD331958 has properties on SIMBAD listed B-V=+1.19 mag, consistent with a K5 star not the M4 spectrum we see. SIMBAD lists its spectral type as B8, which is inconsistent with both K5 and M4. There are multiple inconsistencies. Its parallax is 14.16 mas and proper motion is 78.5 mas/yr, implying a transverse space velocity of 26 km s$^{-1}$, which is consistent with a Milky Way disk star and with its measured radial velocity of 48 km s$^{-1}$. The Vmag and parallax imply an absolute magnitude of $M_V$ = 6.62, which is consistent with K dwarf implied by the B-V. However, the spectrum we see is clearly an M4-M5 dwarf from the obvious TiO bands. Thus, HD 331958 seems unlikely to be the candidate shown in Figure 13.

The other nearby catalog star of comparable brightness is TYC 2675-1608-1 (20 13 10, DEC=+33 07) with SIMBAD photometry, V=11.86 and B-V=+1.65 and J-V = +3.28, sufficiently red to be an M4 star. Its listed parallax is 0.623 mas, implying a distance of 1600 pc. That great distance rules out M dwarf status for a star having Vmag=11.86. One solution that satisfies the parallax, Vmag, and the observed M4 spectral type is an M supergiant at 1600 pc having $M_V$ ~ 0. The measured proper motion of 12 mas yr$^{-1}$ and distance ~1600 pc implies a transverse velocity of ~91 km s$^{-1}$, much larger than the stated radial velocity of 0.12 km s$^{-1}$ listed on SIMBAD. Such a mismatch of velocity components raises concerns about a mistake somewhere. In any case, the apparent emission appears to be noise. Follow-up spectra are certainly warranted to verify that the apparent emission at 827 nm is indeed merely noise.

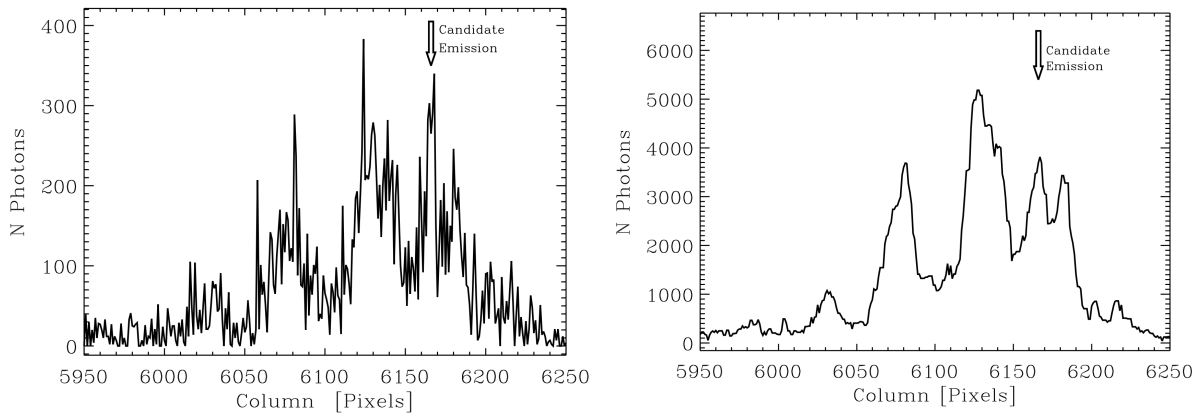

*Figure 13. A candidate transient emission line in an M dwarf at Galactic Longitude 72 deg. At left: The extracted spectrum vs. Column #, showing the location of the candidate emission found by the automated code. At right: The sum of 20 exposures, showing that the candidate emission (top) resides where there is persistent emission, but is consistent with the noise in a single spectrum, given 600 exposures to draw from, as shown in Figure 14.*

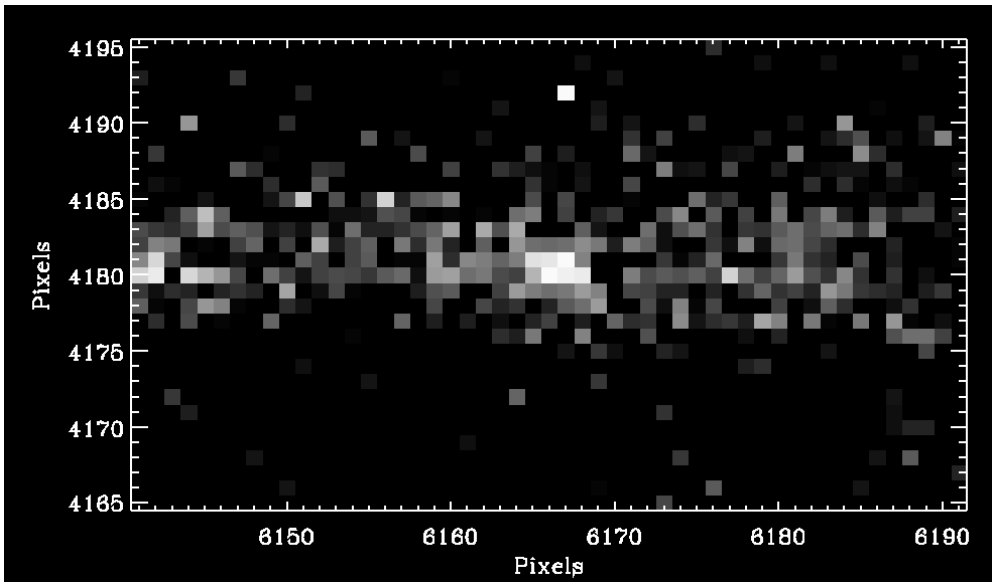

*Figure 14. The raw image of the emission candidate shown in Figure 11. The bright pixels near the center are not distributed smoothly over the full vertical length of the PSF in the spatial direction. This indicates the brightening of "emission" is due to noise or a cosmic ray.*

At Galactic longitude 64 deg the automated difference-image code identified candidate transient emission located on the spectrum of a star. Figure 15 shows a 1.2x1.2 deg zoom with the spectrum of the star in the NE corner. The spectrum is short in wavelength, and it exhibits four broad, bright wavelength domains, typical of the molecular bands of M stars with most of the flux longward of 700 nm. Identification and astrometry of five stars in the vicinity, shown in Fig. 15, shows that the mid-M-type star is at RA = $19^h\ 45^m\ 38.0^s$ and DEC=+28° 39' 40" (2000), with an uncertainty of 2 arcmin, and it is magnitude R~11. The large astrometric and photometric uncertainty is due to the dispersion by the prism in DEC and to the vignetting in the corner of the image.

The transient emission appears in only one exposure (#588 out of 600) and is located on a broad peak at 745 nm as shown in Figure 16. The stellar spectrum indeed exhibits four broad peaks in the spectral energy distribution due to the usual absorption by TiO, CaH, CaOH in mid-M dwarfs. Figure 16 shows the stellar spectrum between wavelengths 710 and 920 nm. There is very little stellar flux detectable outside that wavelength range. The enhanced emission at 745 nm is apparent in the 1-sec exposure (shown as squares) relative to the average spectrum (solid line) from 20 exposures, with 10 taken before and 10 after.

In the right panel of Figure 16, the individual 20 spectra are shown that comprised the average. The scatter in the number of photons at any given wavelength reveals the noise in the individual 1-sec exposures. The noise is caused by Poisson fluctuations of photon arrival and seeing variations. The dispersion of the number of photons, at each wavelength, among the 20 exposures shows the level of combined noise from Poisson fluctuations and seeing. The enhanced emission in the one exposure at 745 nm triggered the automated search algorithm (squares), and it is indeed more intense than the average. However, that enhanced intensity has a magnitude that resides at the end of a distribution of noise rather than detached from that distribution. Thus, the apparent transient emission is likely a result of rare, but expected, noise as seen in the other spectra. Further, stellar photospheric flux at 745 nm is naturally the most intense region of the spectrum, which allows a slight seeing improvement to concentrate the arriving photons in both the wavelength and spatial directions of the raw image, boosting the peak intensity momentarily. We are satisfied that this apparent transient emission is merely noise.

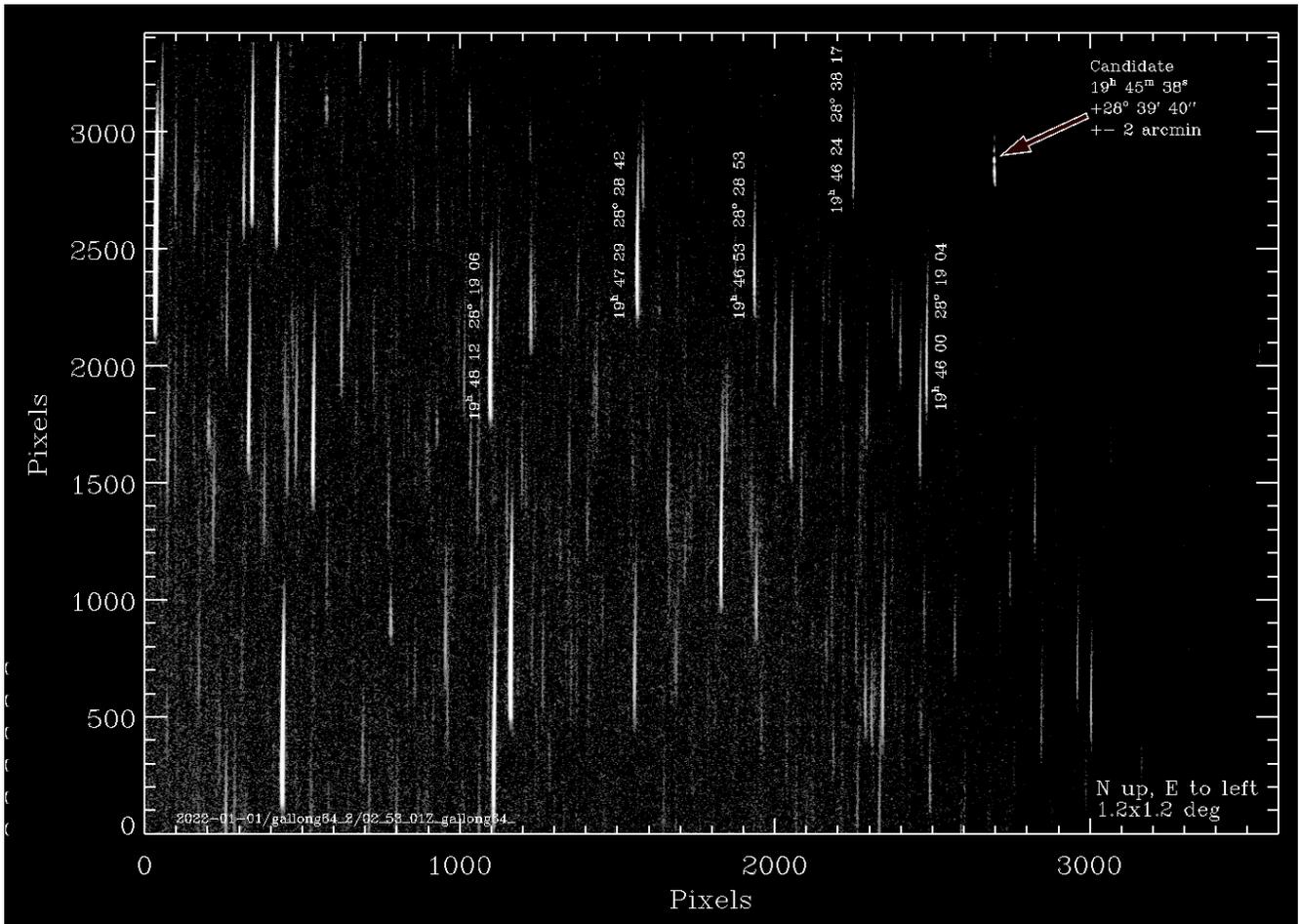

*Figure 15. The sum of 20 consecutive 1-sec images, zoomed on a transient emission-line candidate at Galactic longitude 64 deg, at upper right. Spectra of stars of brightness Vmag 7 to 14 are visible, with North up, short wavelengths up, and East to the left. The stellar spectrum with the emission candidate is short because it emits nearly all of its light in the red and infrared. The coordinates of five identified stars are shown, but the identity of the red star with the candidate emission remains unidentified. The transient emission is shown in Figure 16, and it is consistent with noise fluctuations from seeing changes and Poisson noise.*

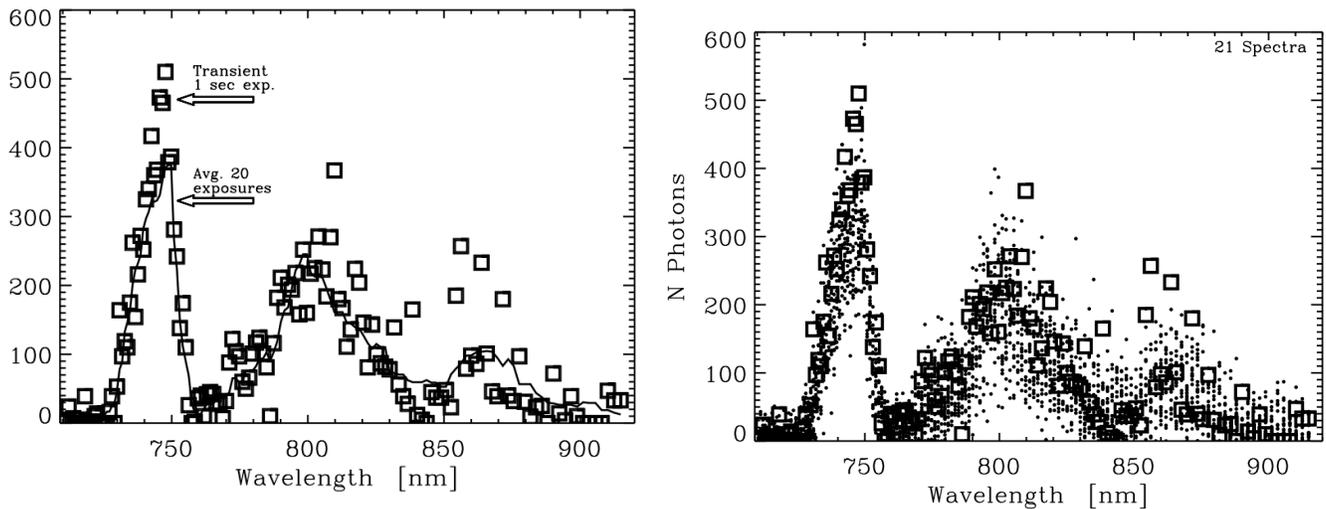

*Figure 16. Spectrum of the apparent transient emission-line at 745 nm in a 1-sec exposure (squares). Left: The solid line is the average of 20 exposures surrounding the one exposure with the enhanced emission. Note the enhanced emission at 745 nm. Right: The 20 individual spectra are shown as small dots, conveying the scatter in the number of photons in each exposure. The apparent transient emission at 745 nm indeed stands more intense than the ensemble of individual spectra. But the scatter in the number of photons at each wavelength shows that the enhanced emission is at the extreme end of that distribution, but not detached from that distribution. Thus the enhanced emission flagged by the automatic detection algorithm is justified, but not inconsistent with the end of the distribution of noise.*

We could not identify a definite M-type star near the coordinates above. The two reddest stars within the 3 arcmin error circle are IRAS 19436+2834 (19 45 40.4 +28 41 55) that has V=10.72, J=5.25, and K=3.733, and IRAS 19433 +2829 (19 45 24, +28 36 55) that has magnitudes G=13.8, J=7.40, and K=5.44. Both stars have colors consistent with a mid to late M-type star as observed here. The first of them is closer to the coordinates measured here. We have no other candidate stars. In any case, as noted, the apparent emission at 745 nm is most likely mere noise from photon-arrival statistics and seeing variations.

At Galactic longitude 248 deg, the automatic algorithm identified a similar apparent transient emission in an M dwarf at RA=$8^h$ $01^m$ $26^s$ DEC=$-30°$ 15' 11". The star is magnitude, R ~ 10, with the continuum heavily chopped by TiO absorption, typical for a star of spectral type ~M3. Figure 17 shows, at left, the spectrum from the 1-sec exposure, and at right, the average spectrum of 20 exposures. There is apparent enhanced emission at 720 nm that is likely to be due to momentary photon-arrival fluctuations and improved seeing.

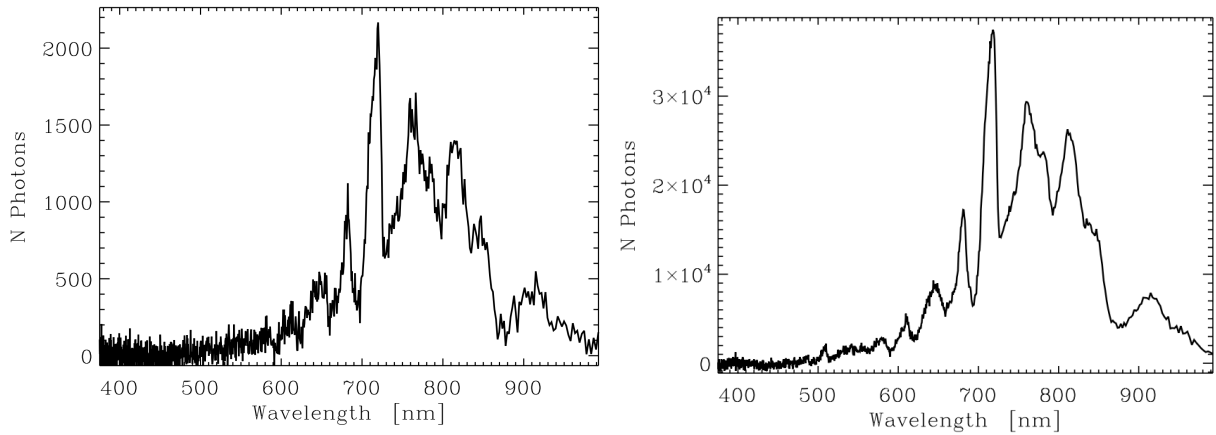

*Figure 17. Another example of an M dwarf spectrum that triggered the automatic search algorithm for transient emission, this being at 720 nm where a natural peak in the stellar spectrum occurs. Momentary excellent seeing peaks up the natural peak in the stellar spectrum.*

### 4.4 P Cygni

The automated difference-image search for transient monochromatic light found "transient" line emission at Galactic longitude 76 deg on 2022 Dec 02. The variable line emission appeared in multiple exposures, suggesting that it was actually constant emission fluctuating due to seeing variations. Figure 18 shows the entire spectrum of this candidate, revealing that the emission line has a wavelength of 656 nm, consistent with H-$\alpha$. Seeing variations no doubt caused the H-$\alpha$ intensity to vary by ~10%, occasionally triggering a "detection" by the difference-image algorithm. The spectrum also has H-$\beta$ in emission and the telluric absorption at the A-band and B-band. There is also emission apparently at 588 nm and 503 nm, likely from HeI. The star has Vmag ~ 5 and its approximate coordinates are RA = 20h 24 m and DEC = +37d 30', consistent with the well-known iconic star, P Cygni.

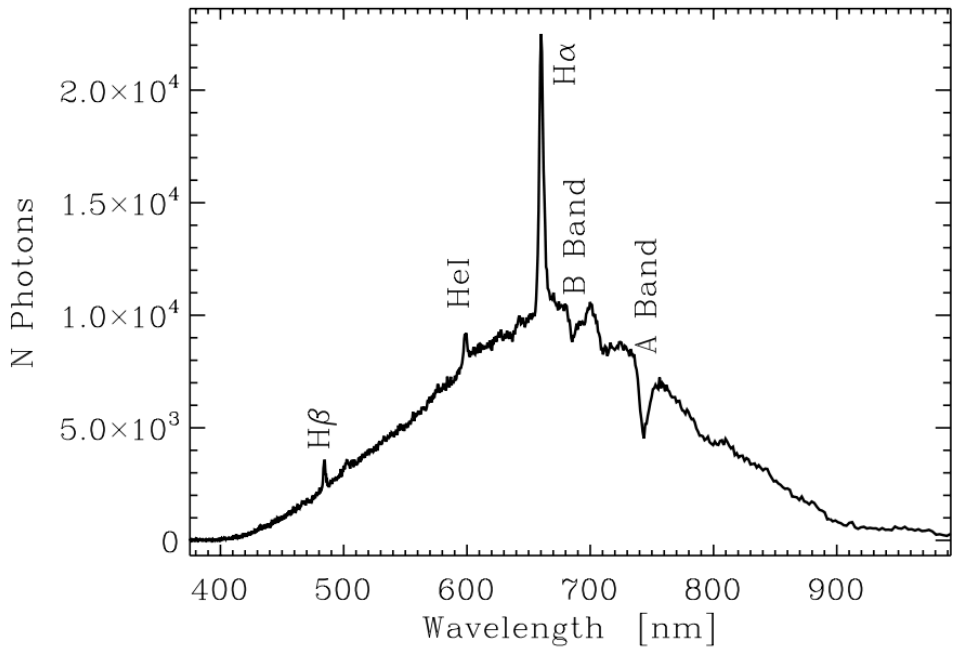

*Figure 18. A 1-sec exposure revealing variable strong emission line, found by the automated code, that triggered a detailed examination by eye. The wavelength scale shows the strong emission line is H-α, and the spectrum also contains emission at H-β and HeI* **(587.6 nm),** *and fainter emission lines. Coordinates show this object is probably P Cygni itself, with spectral resolution inadequate to reveal the blueward absorption, but showing a steep shortward edge of its profile.*

### 4.5 Be Star or Planetary Nebula

At Galactic longitude 112 the automatic search algorithm identified a transient emission line at RA ~ $23^h 24^m 51^s$ DEC~$+61°$ 14 18" (within 30"), based on astrometry calibrated by neighboring stars. The extracted spectrum is shown in Figure 19, the sum of 20 1-s exposures. Strong Hα triggered the difference-image search algorithm due, no doubt, to seeing changes. The object has a continuum that is blue, similar to stars of spectral type early A or B-type. The Balmer emission lines have a sharp blueward edge indicative of gas outflow, and the spectrum contains other emission lines common from $10^4$ K gas. Lists of stars within a 1 arcmin revealed no obvious identifications, with closest being an A2III star BD+60 2536 (V=9.57, B=9.93), with no spectrum published to check for emission lines. The blue continuum and emission line profiles suggest an early type star with mass outflow, somewhat reminiscent of KjPn8 (Vazquez, Kingsburgh, and Lopez 1998). In any case, the spectrum is consistent with an astrophysical explanation, removing it relevant for our purposes. It deserves follow-up spectroscopy.

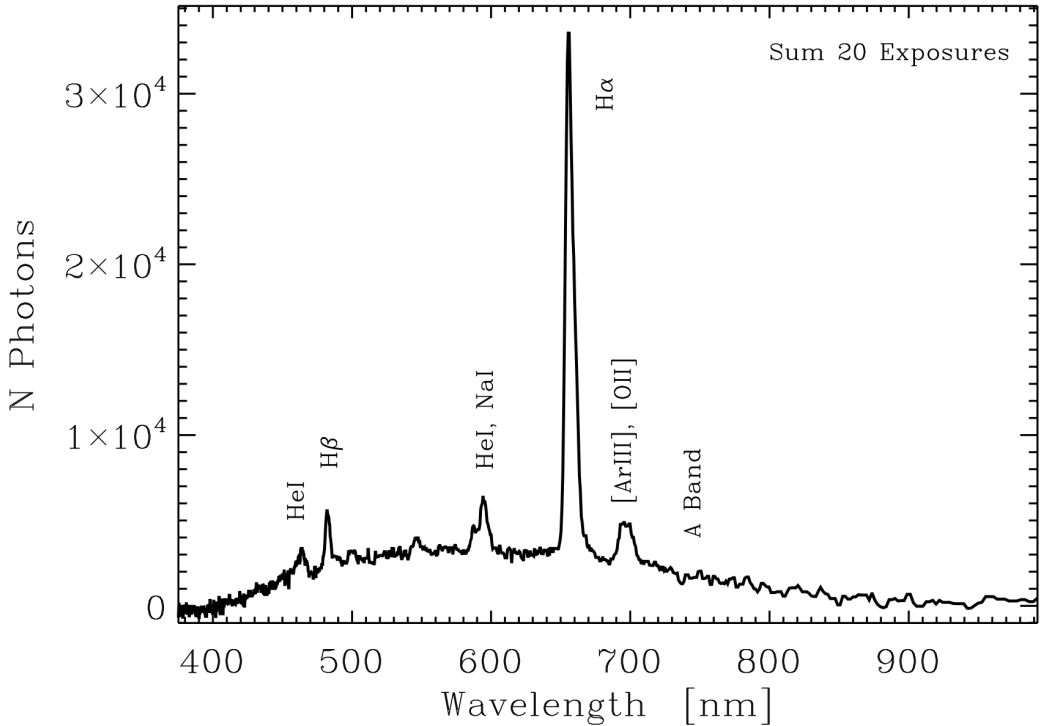

*Figure 19. A candidate transient emission candidate at Galactic longitude 112 deg that is simply an astrophysical object with Hα that varied due to seeing as is common. The spectrum is likely an V ~ 11 mag Be star or planetary nebula (or both), and the triggering transient emission is at the wavelength of Hα. This blue object at RA ~ $23^h\ 24^m\ 51^s$ DEC~$+61°\ 14\ 18"$ (within 30") remains unidentified.*

### 4.6 Summary of Monochromatic Candidates

The difference-image algorithm performed a search of 124 fields, each 3.2x2.1 deg, along the Galactic Plane in this objective prism survey. The limiting magnitude for monochromatic emission was approximately Vmag = 13. Added to the previously observed fields near the Galactic Centre, a total of 973 square degrees were surveyed near the entire Galactic Plane accessible from latitude 38 deg. Each field was observed with 600 consecutive 1-sec exposures, allowing a difference-image search for monochromatic objects of interest. The automated difference-image search for transient emission identified 11 candidate sources of monochromatic emission, some of which were continuously emitting but fluctuating due to seeing variations. We carefully examined each candidate, including any associated stellar spectrum. None of the 11 candidates were pulses nor continuous emission of monochromatic light. Instead, all of them were either astrophysical objects with a strong emission feature in the spectrum that varied due to seeing changes or aircraft with flashing xenon lights, including one that was nearly monochromatic. Unidentified aircraft or spacecraft that have unexpected spectral characteristics would stand out. *We found no point-sources of monochromatic emission, pulsing or continuous, that were plausibly extraterrestrial lasers.*

## 5 DETECTION EFFICIENCY: INJECTION AND RECOVERY OF LASER PULSES

We generated 100 synthetic monochromatic pulses consisting of 2D Gaussians having FWHM ~ 6 pixels, representative of the actual PSF of our images of the Milky Way Plane. We scaled these synthetic monochromatic pulses to various total numbers of photons within the entire profile, from 400 to 1000 photons. These synthetic pulses ranged from roughly 0.2x background to 1.5x the background photons per pixel. We added these synthetic pulses to actual individual images, simulating a pulse duration less than 1 sec. We placed the pulses at random locations within the image, both in between and coincident with stellar spectra.

For each of these real images with injected monochromatic pulses, we executed the blind difference-image analysis to determine if it "discovered" the synthetic pulses. We ran 100 cases for each level of pulse intensity. The fraction of injected pulses detected is shown graphically in Figure 20. Blindly executing the image-difference algorithm described above, we found the code successfully discovered 50% of the injected pulses that had at least 650 total photons in the profile. It found none of the pulses containing fewer than 500 photons, and it found 97% of the pulses having more than 900 photons.

Thus, the nominal detection threshold at which 50% of the pulses would be detected is 650 photons total within the monochromatic pulse. *This 650 photon threshold represents the number of photons that must be detected in 1 sec such that half of such pulses would be detected.* The search algorithm has diminishing sensitivity for pulses lasting over the 1 sec exposure time of each frame. For such cases, some of the adjacent six bookend exposures would contain the emission, diminishing their contrast with the target image. In particular, for continuous monochromatic emission, ~6500 photons per sec would be required in order for the 10% variations to reveal itself as a pseudo-pulse. The term "continuous" here refers to a cadence of pulses that is more frequent than 1 pulse per second. A train of pulses of nanosecond duration and arriving $10^6$ per second would be detected here only as "continuous" monochromatic emission, requiring seeing variations for detection.

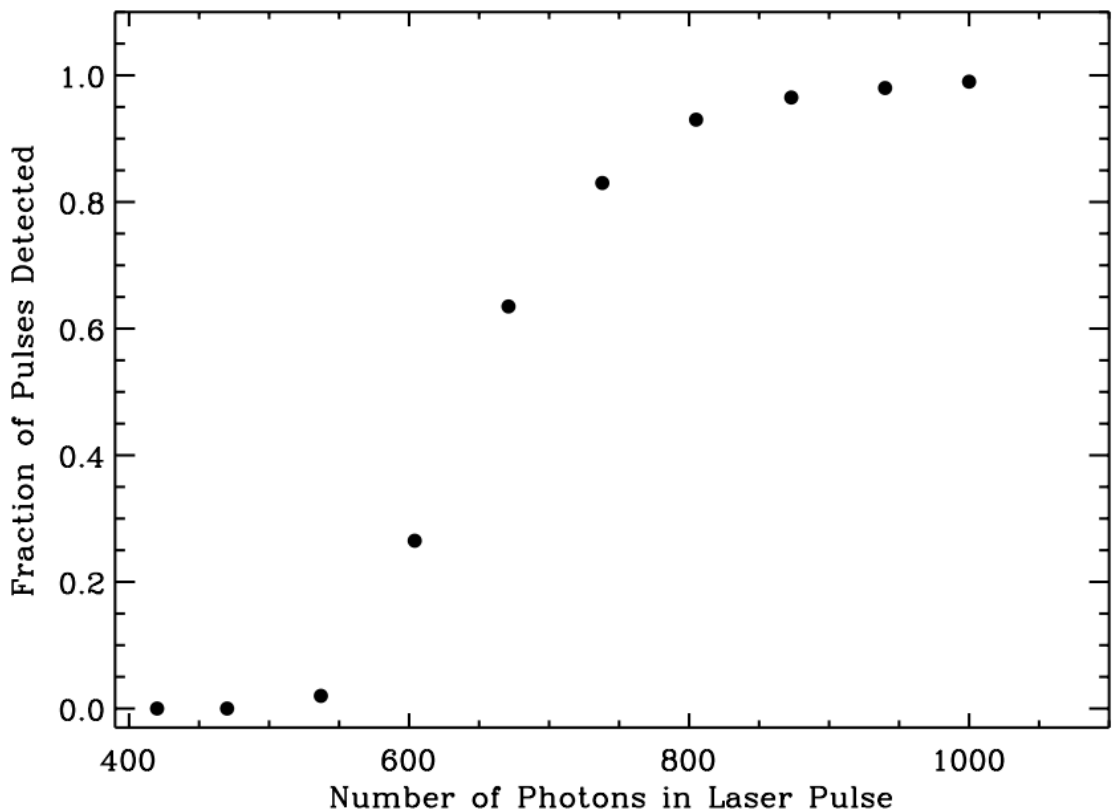

*Figure 20. The fraction of injected monochromatic pulses detected blindly by the difference-image search algorithm as a function of the number of photons in the monochromatic pulse. Pulses containing 650 photons (total within the PSF) are detected in 50% of trials. Pulses with >1000 photons are detected in ~100% of the trials. The nominal detection threshold is 650 photons per laser pulse.*

## 6. DISCUSSION

We searched 2/3 of the Galactic Plane in a swath 2 deg wide for sub-second pulses of monochromatic emission between wavelengths 380 and 950 nm. The technique was also sensitive to sources of constant monochromatic emission as seeing causes momentary 10% fluctuations in the acquired photons in a one-second exposure. The goal was to search a domain of transients, in time and wavelength, that could have been missed in past transient surveys that use exposure times over 30 sec and usually had modest or no spectroscopic ability to detect unexpected emission lines. For example, searches for planetary nebulae, HII regions, and flare stars used exposures more than 1 minute and were often confined to the detection of Balmer lines, especially Hα.

The detection threshold of 650 photons within a pulse translates into a threshold of photon fluxes entering the telescope. The 650-photon threshold corresponds to a fluence per unit area by using the effective collecting area of the 0.278-m RASA telescope system, including efficiency between 450 – 800 nm and blockage by the camera at prime focus. We find that the effective collecting area is $A_{eff}$ = 0.020 m$^2$ (Marcy, Tellis, and Wishnow 2021,2022). Thus, the detection threshold of 650 photons implies a fluence threshold of 32500 photons per square meter at the Earth's surface for monochromatic pulses of duration less than 1 sec. For a pulse lasting 1 sec, that fluence corresponds to Vmag = 15. For wavelengths below 450 nm and above 800 nm the quantum efficiency drops below 50% of peak QE (at ~600 nm), thus requiring more than 32500 photons per square meter for detection. Atmospheric extinction raises this threshold fluence at the top of the Earth's atmosphere by a few percent.

One driver for this new search was the detection of laser beams in the Galaxy. A monochromatic light source lasting a few nanoseconds, microseconds, or milliseconds would have been detected in one exposure relative to reference exposures, with a detection threshold of 650 photons in the pulse. *We found no pulsed monochromatic sources, nor any unknown continuous monochromatic sources, between Galactic Longitude -4 to 248 deg, in a swath ~2 deg wide along the Galactic Plane.*

A major consideration in this optical SETI program was to minimize false positives. We engineered the optics, pixel size, and difference-imaging algorithm thresholds to avoid false positives, such as from cosmic rays, satellite glints, Cherenkov radiation, or electronics noise. Our entire system, end-to-end, was designed to avoid them, and indeed we found none, after careful scrutiny of candidates. Without a doubt, the optics, detector, and algorithm could be modified to make the entire system more sensitive to monochromatic pulses, by allowing some false positives to pass through. But we elected to choose parameters that avoided the difficult task of identifying them (e.g., Sheikh et al. 2021, Villarroel et al. 2022).

Some pulse cadences would not have been detected. For example, a duty cycle of one pulse per day, week, or month, implies that one pulse could arrive while we were not observing. Obviously, arbitrary pulse cadences with low duty cycle bring a lower detection probability. More observations are warranted to provide greater cadence coverage. This survey effectively searches for cadences of at least one pulse every 10 minutes, the duration of the observing sequence of a given field. Pulse cadences more rapid than 1 Hz are effectively "continuous" in time. In such cases, the detection threshold is approximately 10x higher, as only the ~10% fluctuations from seeing variations will cause such continuous sources to vary above the detection threshold in the difference images.

We can compute the laser power needed, from interstellar distances, to produce a detectable fluence. We consider a benchmark laser that is diffraction limited with a 10-meter diameter aperture, and located 100 ly from Earth. Extinction by dust is ~10%. It emits a beam with an opening angle of ~0.01 arcsec at a wavelength of, say, 500 nm. To produce a photon flux at Earth having the detection threshold of 32500 photons per square meter, a power of 100 Megawatt is required during the 1 sec pulse. For a laser launcher located 1000 ly away, a power of 10 Gigawatt is required. Extinction from interstellar dust will increase this requirement by ~30%. At 1 kpc, the required power is ~90 Gigawatt. Extinction from dust will require another factor of two in power. The laser beam footprint at Earth would have a diameter of 0.3 to 3 AU, respectively, a fraction of the area of the inner Solar System. Such small footprints imply that the laser is purposely or fortuitously (or mistakenly) pointed at Earth. Lasers having apertures smaller than 10 meter can also be detected, but their larger opening beam angle would require more laser power, increasing inversely as the square of aperture diameter.

A benchmark receiver might have 100x the diameter of the collector and thus 10,000x higher collecting area. This larger collector can thus reliably receive 10,000x the bit rate here (Gertz & Marcy 2022). A string of cell towers overcomes this limitation of bit-rate.

## 7. SUMMARY

More than 5000 stars have been searched for monochromatic laser emission (Reines and Marcy 2002, Tellis and Marcy 2015, 2017, Marcy 2021, and Marcy et al. 2022). Most surveyed stars are Sun-like or smaller with spectral types F, G, K, or M-type. But several hundred are massive stars of spectral types O, B, A, and early F-type (Tellis, private communication 2022). We have also searched the focal points of the solar gravitational lens (Marcy et al. 2021). Many searches for optical pulses of sub-second duration have been performed, covering over half the sky (e.g. Wright et al. 2001, Stone et al. 2005, Howard et al. 2004, Howard et al. 2007, Maire et al. 2020). *No optical laser emission, pulsed, monochromatic, or otherwise, has ever been detected.*

Optical lasers could have been detected previously by "conventional" astrophysics searches, both broad-band and spectroscopic. Objective prism telescopes 100 years ago (Fleming et al. 1907, Pickering 1912, Cannon & Pickering 1922) revealed thousands of objects that emit emission lines, such as planetary nebulae, HII regions, T Tauri stars, Be stars, Wolf-Rayet stars, M dwarf flare stars, and active galactic nuclei, including at high redshift. Such objective prism surveys yield emission-line objects that are then pursued with spectroscopy of modest resolution that is able to reveal non-astrophysical emission lines if any existed. Even optical searches using mere broadband filters, such as BVRI, would reveal objects whose emission was confined to one or two bands. A 13$^{th}$ mag source whose emission was confined to one or two emission lines would exhibit bizarre colors, meriting follow-up spectroscopy (ala Lyman-break galaxies). The emission at non-astrophysical wavelength would be immediately obvious. Many all-sky surveys reached 18$^{th}$ magnitude, and modern ones reach 21$^{st}$ magnitude. No object with strange colors turned out to be lasers. ***In summary no monochromatic sources with non-astrophysical emission lines, e.g., lasers, were discovered in hundreds of past all-sky surveys, nor in the present survey.***

The present non-detection of lasers, and those from the hundreds of past surveys, does not imply that extraterrestrial technology is absent in the Milky Way. To be sure, other wavelengths and pulse cadences merit observations. Also, the filling factor of optical laser beams may be too small (Forgan 2014). A vast parameter space is yet to be surveyed (Wright et al. 2018). However, a large fraction of *observable SETI parameter space* has already been surveyed by both conventional astronomy surveys of the entire sky and by explicit searches for extraterrestrial technology (e.g., Wlodarczyk-Sroka, Garrett & Siemion 2021, Garrett & Siemion 2022).

Most impressively, SETI parameter space has *unintentionally* been surveyed by all-sky searches for natural, astrophysical objects. Those surveys gloriously revealed a multitude of astrophysical objects that emit, unexpectedly, at all wavelengths, including radio, microwave, extreme UV, x-rays, and gamma rays. Those unexpected discoveries constitute *non-detections* of extraterrestrial technology in the same search domains, but not recorded as such. The dearth of beacons of technology leaves us revealing more of a great SETI desert, especially at the intensely surveyed optical and radio wavelengths.

## ACKNOWLEDGMENTS

This work benefitted from valuable communications with Beatriz Villarroel, Franklin Antonio, John Gertz, Ben Zuckerman, Brian Hill, Susan Kegley, Dan Werthimer, Ariana Paul, Roger Bland, Martin Ward, and Paul Horowitz. We thank the team at Space Laser Awareness for outstanding technical help.

## DATA AVAILABILITY

This paper is based on raw CMOS sensor images obtained with Space Laser Awareness double objective prism telescopes. The 8864 raw images are 125 MB each, totaling 1.1 TB. They are located on a peripheral disk that is not online. All images are available to the public upon the request of G.M., and a transfer method must be identified.

This paper was typeset from Microsoft WORD document prepared by the author.